\documentclass[twocolumn, pra, aps, superscriptaddress, longbibliography,showpacs,amsmath,amssymb,floatfix
]{revtex4-1}

\usepackage{graphicx}
\usepackage{dcolumn}
\usepackage{bm}
\usepackage{graphicx}             \usepackage{booktabs}                                  
\usepackage{amssymb}

\usepackage{amsmath}
\usepackage{epsfig}
\usepackage{xcolor}
\usepackage{tabu}
\usepackage{mathtools}
\usepackage[colorlinks,linkcolor=blue,anchorcolor=blue,citecolor=blue,urlcolor=blue]{hyperref}
\usepackage{physics}
\usepackage{float}
\usepackage{diagbox}
\usepackage{inputenc}

\begin{document}

\title{On the different Floquet Hamiltonians in a periodic-driven Bose-Josephson junction} 
\author{Xiaoshui Lin}
\email{lxsphys@mail.ustc.edu.cn}
\affiliation{CAS Key Laboratory of Quantum Information, University of Science and Technology of China, Hefei, 230026, China}
\author{Zeyu Rao}
\affiliation{CAS Key Laboratory of Quantum Information, University of Science and Technology of China, Hefei, 230026, China}
\author{Ming Gong}
\email{gongm@ustc.edu.cn}
\affiliation{CAS Key Laboratory of Quantum Information, University of Science and Technology of China, Hefei, 230026, China}
\affiliation{Synergetic Innovation Center of Quantum Information and Quantum Physics, University of Science and Technology of China, Hefei, Anhui 230026, China}
\affiliation{Hefei National Laboratory, University of Science and Technology of China, Hefei 230088, China}

\date{\today}

\begin{abstract}
The bosonic Josephson junction, one of the maximally simple models for periodic-driven many-body systems, has been intensively studied in the past two decades. 
Here, we revisit this problem with five different methods, all of which have solid theoretical reasoning.
We find that to the order of $\omega^{-2}$ ($\omega$ is the modulating frequency), these approaches will yield slightly different Floquet Hamiltonians.
In particular, the parameters in the Floquet Hamiltonians may be unchanged, increased, or decreased, depending on the approximations used. 
Especially, some of the methods generate new interactions, which still preserve the total number of particles; and the others do not.
The validity of these five effective models is verified using dynamics of population imbalance and self-trapping phase transition.
In all results, we find the method by first performing a unitary rotation to the Hamiltonian will have the highest accuracy.
The difference between them will become significate when the modulating frequency is comparable with the driving amplitude.
The results presented in this work indicate that the analysis of the Floquet Hamiltonian has some kind of subjectivity, which will become an important issue in future experiments with the increasing of precision.
We demonstrate this physics using a Bose-Josephson junction, and it is to be hoped that the validity of these methods and their tiny differences put forward in this work can be verified in realistic experiments in future using quantum simulating platforms, including but not limited to ultracold atoms. 
\end{abstract}

\maketitle

\section{Introduction}

Since the first observation of Bose-Einstein condensate (BEC) in ultracold atom gas \cite{Anderson1995Observation, Davis1995Bose}, ultracold atom has become one of the most active research topics in the past two decades \cite{Leggett2001BoseEinstein, Dalfovo1999Theory, Cornell2002Nobel, Ueda2010Fundamentals}.
Due to its long coherent time and the flexible tunability of interactions and external potentials, this system offers a perfect platform for researching macroscopic quantum phenomena, including interference of matter-wave \cite{Andrews1997Observation}, quantized vortex lattice \cite{Fetter2001Vortices}, Josephson oscillation and self-trapping effects \cite{Albiez2005DirectObservation, Levy2007AcDcJosephson}, to name only a few of them. 
Furthermore, with Feshbach resonance \cite{Chin2010Feshbach}, the scattering length of the bosonic atomic gas can be well-tuned, which opens an avenue for researching physics from the weakly correlated physics to the strongly correlated physics with these ultracold gases \cite{Ho2004Universal, Randeria2014Crossover, Makotyn2014Universal, Cowell2002Cold, Eigen2017Universal}. This physics can also be explored with ultracold Fermi gases. 

\begin{figure}[htbp!]
\centering
\includegraphics[width=0.32\textwidth]{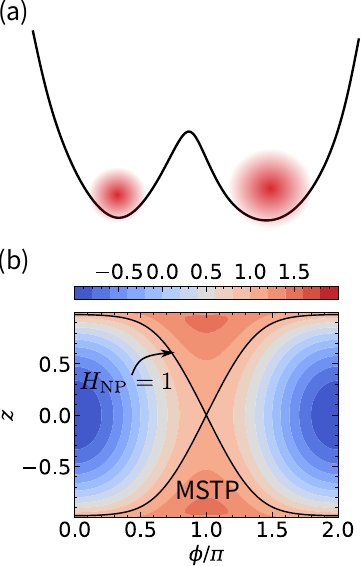}
\caption{(a) The schematic of the bosonic Josephson junction, which is generated by confining the BEC in a double well. (b) The constant energy surface of the nonrigid pendulum at $\Delta = 0$ and $\nu = 1$. When $H_{\text{NP}}>\nu = 1$, the system is in the MSTP with $z(t) \neq 0$ for any time.}
\label{fig-schematic}
\end{figure}

In these researches, the bosonic Josephson junction [see Fig. \ref{fig-schematic} (a)] plays a unique role because of its simplicity and tunability.
This model can be written in the second quantization representation as 
\begin{equation}
    H = \int dx\hat{\Psi}^{\dagger}(x)[-\frac{\hbar^2}{2m}\nabla^2 + V(x) + \frac{g}{2}\hat{\Psi}^{\dagger}(x)\hat{\Psi}(x)]\hat{\Psi}(x),
\end{equation}
where $V(x)$ is the double well potential, $m$ is the atom mass, and $g = 4\pi\hbar^2 a_s/m $ with $a_s$ the scattering length and $\hbar$ is the reduced Planck constant.
The scattering length can be tuned via Feshbach resonance \cite{Chin2010Feshbach}. 
When the potential depth is large, the lowest two energy levels are well separated from the other excited energy levels.
In the two modes approximation, we can approximate the field operator using $\hat{\Psi}(x) = \hat{a} \psi_\text{L}(x) + \hat{b} \psi_\text{R}(x)$, where $\psi_\text{L}(x)$ and $\psi_\text{R}(x)$ are the ground state wave functions of each well under harmonic approximation \cite{Milburn1997Quantum}.
The overlap between wave functions $\psi_\text{L}(x)$ and $\psi_{R}(x)$ is small and its strength is characterized by $\nu$.
Furthermore, the detuning between the two wells is given by $\Delta$, and the on-site interaction strength, assumed to be the same, is given by $\lambda$. 
hence, we should have the following effective Hamiltonian
\begin{equation}
H = \Delta(\hat{a}^\dagger \hat{a} - \hat{b}^\dagger \hat{b}) - \nu (\hat{a}^\dagger \hat{b}+ \hat{b}^\dagger \hat{a}) + \lambda (\hat{a}^{\dagger}\hat{a}^{\dagger} \hat{a} \hat{a} + \hat{b}^{\dagger}\hat{b}^{\dagger} \hat{b} \hat{b} ),
\label{eq-josephson-junction-orgin}
\end{equation}
where $\hat{a}$ and $\hat{b}$ are the annihilation bosonic operators in each well. 
By changing the parameters $\Delta$, $\nu$, and $\lambda$, this model can undergo a transition from the oscillating phase to the macroscopic self-trapping phase (MSTP) \cite{Raghavan1999Coherent, Smerzi1997Quantum, Milburn1997Quantum}.
We can use a mean field approximation to understand this transition by replacing the operators $\hat{a}$ and $\hat{b}$ with some complex numbers $a$ and $b$, which is expected to be valid in the large $N$ limit.
In this way, $\hat{a}^\dagger$ and $\hat{b}^\dagger$ are approximated by $a^*$ and $b^*$, respectively. 
Then we can map the mean field Hamiltonian to a nonrigid pendulum \cite{Smerzi1997Quantum} in phase space with 
\begin{equation}
H_{\text{NP}} = \frac{\lambda}{2}z^2 + \Delta z - \nu \sqrt{1 - z^2}\cos(\phi),
\label{eq-nonrigid-pendulum}
\end{equation}
where $z = |a|^2 - |b|^2$ and $2 \cos(\phi) = (a^{*}b + b^{*}a)/|ab|$.
The $\phi$ and $z$ are conjugate coordinates and momentum, respectively.
The equation of motion is given by 
\begin{equation}
\dot{\phi} = {\partial H_\text{NP} \over \partial z}, \quad 
\dot{z} = - {\partial H_\text{NP} \over \partial \phi}.
\end{equation} 
Furthermore, with energy conservation, it is found that the value of $z(t) = 0$ becomes inaccessible at any time if 
\begin{equation}
    \lambda > 2 \nu \frac{\sqrt{1 - z(0)^2}\cos(\phi(0)) + 1}{z(0)^2}, 
    \label{eq-self-trapping-criteria}
\end{equation}
for $\Delta = 0$ [see Fig. \ref{fig-schematic} (b)]. Especially, we find $\lambda_c = 2\nu$ for $z(0) = 1$ and $\phi(0) = 0$.
This criterion for the MSTP works well for the mean field dynamics. 
A detailed analysis of this problem in mean field theory in terms of Josephson effect, $\pi$ oscillations, symmetry breaking, and macroscopic quantum self-trapping have been presented in Ref. \cite{Raghavan1999Coherent}. 
Intriguing generalization of this models to various systems include dipole gases in Ref. \cite{Xiong2009Symmetry}, chaos with single impurity in Ref. \cite{Chen2021Impurity}, exciton-polaritons in Ref. \cite{Abbarchi2013Macroscopic}, Josephson oscillation in triple-well potentials in Ref. \cite{Lahaye2010Mesoscopic}, dynamical stability for BEC in a periodic potential in Ref. \cite{Lellouch2017Parametric}, two coupled one-dimensional condensates in Ref. \cite{Ji2022Floquet}, dissipative Josephson junctions coupling to a bosonic bath in Refs. \cite{Saha2023Phase, Masuki2022Absence}, and spin-orbit coupled condensate in a single well in Ref. \cite{WeifenZhuang2021Angular}, to only mention a few of them. 
The self-trapping and tunneling in a double well potential have been observed in experiments \cite{Albiez2005DirectObservation}.
It can also be found in an array of coupled condensates \cite{Reinhard2013Selftrapping}. 
This physics can also be explored using the other quantum systems, including polaritons, superconductors, and trapped ions {\it etc.}, which are not discussed here. 

\begin{table*}[htbp!]
\centering
\renewcommand\arraystretch{2.5}
\begin{tabular}{|p{0.1\textwidth}|p{0.13\textwidth}|p{0.13\textwidth}|p{0.13\textwidth}|p{0.13\textwidth}|p{0.16\textwidth}|}
       \hline
      Method   & $\Delta' /\Delta $ & $\lambda' / \lambda $ & $\nu' / \nu $ & $V_{\text{new}}$ ($\hat{V}_{\text{new}}$) & $\lambda_c$ at $|A/\omega| \ll 1$ \\
        \hline  
       Q-I  & 1   &  1 & $1 - \frac{A^2}{\omega^2}$ & - & $2\nu(1 - \frac{A^2}{\omega^2})$\\
       \hline
       Q-II  & $1- \frac{\nu^2}{4\omega^2B} $ & $1- \frac{8\nu^2 B}{\omega^2}$ & $J_0 +  \frac{ 2 \nu^2 D}{\omega^2}$ & Eq. \ref{eq-method-2-new} & $2\nu(1 - \frac{A^2}{\omega^2})$ \\
       \hline
       C-I & 1   & $1 + \frac{A^2}{2\omega^2}$ & $1 - \frac{A^2}{\omega^2}$ & - & $2\nu(1 - \frac{3A^2}{2\omega^2})$\\
       \hline
       C-II  & $1- \frac{4 \nu^2 B}{\omega^2} $ & $1- \frac{6 \nu^2 B}{\omega^2}$ & $J_0  + \frac{2\nu^2 M}{\omega^2}$ & Eq. \ref{eq-method-4-new} & $2\nu(1 - \frac{A^2}{\omega^2})$\\
       \hline
       C-III  & $1 + \frac{A^2}{\omega^2}$ & $1 + \frac{A^2}{\omega^2}$ & $1 - \frac{A^2}{2\omega^2}$ & - & $2\nu(1 - \frac{3A^2}{2\omega^2})$ \\
       \hline
    \end{tabular}
    \caption{Summary of the parameter changes and new terms for the Floquet Hamiltonian under five different methods. Here, the parameters $B$, $C$, $M$, $D$, $\cdots$, can be found in the main text. }
    \label{tab-summary}
\end{table*}

More intriguing physics can be found in this simple model in the presence of periodic driving.
For instance, periodic driving can induce coherent destruction of tunneling \cite{Kayanuma2008Coherent, Gong2009Many-body, Kierig2008Single} for dynamic localization, Shapiro-like resonant effect \cite{Raghavan1999Coherent, Eckardt2005Analog}, as well as the Hamiltonian chaos \cite{Lee2001Chaotic, Weiss2008Differences, Salmond2002Dynamics, Zhang2008Nonlinear, Jiang2014Equilibrium}.
This system with periodic modulation has been studied intensively in Refs. \cite{Holthaus2001Coherent, Holthaus2001Towards, Xie2007Nonlinear, Xie2009Nonlinear, Abdullaev2000Coherent,Luo2008Quasienergies,Watanabe2012Floquet, Abdullaev2000Coherent}. 
In all this research, the Floquet analysis plays a unique role, and we expect a time-independent Floquet Hamiltonian $\hat{H}_\text{F}$ to determine the stroboscopic dynamics. 
Then many aspects of the periodic-driven systems can be understood with the help of this Floquet Hamiltonian.
However, it is very hard for us to find the exact expression of the Floquet Hamiltonian in the most generic conditions.
This Floquet Hamiltonian is also not unique. 
From different starting points or different approximations, one may yield different effective Hamiltonians. In this way, some of the effective Hamiltonians may work well, and some of them may not. 
Thus, it is crucial for us to identify which method can give the most precise prediction.
To date, this issue has not yet drawn much attention in theory and has not been a concern in experiments, however, it will become a critical issue in the future with the increasing of experimental precision, and it is to be hoped that the different effective models can be verified in these experiments.

\subsection{Physical model}

To address this problem, we consider the following maximally simple model \cite{ Milburn1997Quantum,
Ruostekoski1998BoseEinstein, Stroescu2016Dissipative, Kierig2008Single, Shin2004Atom} (set $\hbar = 1$)
\begin{equation}
\begin{aligned}
    \hat{H} &= (\Delta + A \cos(\omega t)) (\hat{n}_a - \hat{n}_b) - \nu (\hat{a}^{\dagger}\hat{b} + \hat{b}^{\dagger}\hat{a})  \\ 
    &+\frac{\lambda}{2}(\hat{n}_a - \hat{n}_b)^2 - \frac{\lambda}{2}(\hat{n}_a + \hat{n}_b) .
    \end{aligned}
    \label{eq-quantum-BEC}
\end{equation}
with the detuning being periodically modulated, and $\hat{n}_a = \hat{a}^{\dagger}\hat{a} $ and $\hat{n}_b = \hat{b}^{\dagger}\hat{b} $ the number operator in each well. 
When $|\omega|$ is much larger than the magnitude of other parameters, we can neglect the periodic driving, and the time-averaged Hamiltonian is equivalent to Eq. \ref{eq-josephson-junction-orgin}.
It will become important if $\omega$ and all the other parameters are comparable. 
A widely used transformation to this model is $\hat{U} = \exp(-i \frac{A}{\omega}\sin(\omega t )(\hat{a}^{\dagger}\hat{a} - \hat{b}^{\dagger}\hat{b}))$, which gives 
\begin{equation}
\begin{aligned}
    \hat{H}' &= \hat{U}^{\dagger} \hat{H} \hat{U} - i \hat{U}^{\dagger} \partial_t \hat{U} \\ 
    & = \Delta( \hat{n}_a - \hat{n}_b) - (\nu e^{i \frac{2A}{\omega}\sin(\omega t) } \hat{a}^{\dagger}\hat{b} + \mathrm{H.c.}) \\ 
    &+\frac{\lambda}{2}(\hat{n}_a - \hat{n}_b)^2 - \frac{\lambda}{2}(\hat{n}_a + \hat{n}_b),
\end{aligned}
\label{eq-quantum-BEC-tranform}
\end{equation}
with the coupling now being periodically modulated.
Although it is expected that the physics of Eqs. \ref{eq-quantum-BEC} and \ref{eq-quantum-BEC-tranform} should be exactly the same, the resulting Floquet Hamiltonians up to a finite truncation are different in practice.

To clarify this difference, we use quantum and classical methods to derive the corresponding Floquet Hamiltonians for Eqs. \ref{eq-quantum-BEC} and \ref{eq-quantum-BEC-tranform}.
For the quantum methods, we treat these Hamiltonians as quantum operators using the approaches developed in Refs. \cite{Rahav2003Effective,Goldman2014Periodically,Golddman2014Light}. 
For the classical methods \cite{Rahav2003Effective}, we directly employ the multiple scale analysis to the nonlinear mean field equation and separate the slow part from the fast oscillating part.
In all these methods, the resulting Floquet Hamiltonian and mean field dynamics are discussed up to the order of $\omega^{-2}$. 
The central results can be given by (classical Hamiltonian)
\begin{align}
    H_\text{F} &= \Delta' (|a|^2 - |b|^2) - \nu' (a^{*}b + b^{*}a)  +\frac{\lambda'}{2}(|a|^2 - |b|^2)^2 \nonumber \\
    &+ V_{\text{new}}, \label{eq-Floquet-mean-field}
    \end{align}
with $a$, $b \in \mathbb{C}$.
The corresponding quantum Floquet Hamiltonian reads as 
\begin{align}
    \hat{H}_\text{F} &= \Delta' (\hat{n}_a - \hat{n}_b) - \nu' (
    \hat{a}^\dagger \hat{b} + \text{h.c.})  + 
    \frac{\lambda'}{2}( \hat{n}_a - \hat{n}_b)^2 \nonumber \\
    &- \frac{\lambda'}{2}( \hat{n}_a + \hat{n}_b) + \hat{V}_{\text{new}}. \label{eq-Floquet-quantum}
\end{align}
The changed parameters from different methods are summarized in Table \ref{tab-summary}. Due to the preservation of the total number of particles, the new interaction only admits terms like $\hat{n}_a \hat{n}_b$ and $ \hat{a}^\dagger \hat{a}^\dagger \hat{b} \hat{b} + \text{h.c.}$; see below. 

This manuscript is organized as follows. 
In Sec. \ref{sec-effective-ham-quantum}, we derive the Floquet Hamiltonian using quantum methods (methods Q-I and Q-II) for Eqs. \ref{eq-quantum-BEC} and  \ref{eq-quantum-BEC-tranform}.
In Sec. \ref{sec-effective-ham-classic}, we employ multiple scale analysis (methods C-I, C-II, and C-III) to the mean field equation of motion for the Floquet Hamiltonian.
In Sec. \ref{sec-comparision}, we present the numerical verification based on the short-time dynamics and criterion of the MSTP.
In Sec. \ref{sec-conclusion}, we conclude and discuss the possible application of our results.
In Appendix \ref{appendix-pendulum}, we discuss the criterion for the MSTP in the presence of a co-tunneling process. 
In Appendix \ref{appendix-exact}, an exact solution of the Floquet Hamiltonian is presented.
We hope these different effective Hamiltonians put forward in this work can be directly verified in future experiments. 

\section{Floquet analysis based on quantum model}
\label{sec-effective-ham-quantum}

In the first two methods, we employ Floquet analysis for the quantum Hamiltonian in Eq. \ref{eq-quantum-BEC} and Eq. \ref{eq-quantum-BEC-tranform}, which we term as method Q-I and method Q-II.
Then we apply the mean field approximation by replacing all operators with complex numbers.
From the Floquet theorem \cite{Rahav2003Effective}, the Floquet eigenstates and eigenvalues can be solved by the following equation
\begin{equation}
 (\hat{H} - i \partial_t) |u_n \rangle = \epsilon_n |u_n \rangle, 
    \label{eq-ham-floquet}
\end{equation}
with $|u_n (t) \rangle = | u(t + T)\rangle$.
Thus, it is expected that there should exist a family of time-independent Floquet Hamiltonians $\hat{H}_\text{F}$, which has the same eigenstates and eigenvalues with $\hat{H} - i \partial_t$.
This Hamiltonian may be obtained from the unitary $\hat{U} = e^{i\hat{K}(t)}$ as 
\begin{equation}
 \hat{G} = \hat{U} \hat{H} \hat{U}^{\dagger} - i \hat{U} \partial_t \hat{U}^{\dagger}. 
    \label{eq-quantum-floquet}
\end{equation}
Once the transformation $\hat{U}$ makes $\hat{G}$ time-independent, we have found the effective Floquet Hamiltonian $\hat{H}_\text{F} = \hat{G}$ exactly.
In practice, we assume $\hat{K}(t)$ and $\hat{H}_\text{F}$ can be expanded in powers of $\omega^{-1}$ \cite{Rahav2003Effective, Goldman2014Periodically,Golddman2014Light} and the resulting Floquet Hamiltonian up to the order of $\omega^{-2}$ can be written as 
\begin{equation}
  \hat{H}_{\text {F }} = \hat{H}_0 + \hat{H}_{\text{F}}^{(1)} + \hat{H}_{\text{F}}^{(2)} + \mathcal{O}(\omega^{-2}), 
  \label{eq-floquet-quantum-expression}
\end{equation}
with 
\begin{align}
\hat{H}_{\text{F}}^{(1)} &=  \sum_{j=1}^{\infty} \frac{1}{\omega j}[\hat{H}_{j}, \hat{H}_{-j}], \\
\hat{H}_{\text{F}}^{(2)} &=  \sum_{j=1}^{\infty} \frac{1}{2 (\omega j)^2}([[\hat{H}_{j}, \hat{H}_{0}], \hat{H}_{-j}]+\text {H.c.}) + \sum_{j, l=1}^{\infty} \frac{1}{3 \omega^2 j l} \nonumber \\
&([\hat{H}_{j},[\hat{H}_{l}, \hat{H}_{-j-l}]]-[\hat{H}_{j},[\hat{H}_{-l}, \hat{H}_{l-j}]] + \text{H.c.}),
\end{align}
for the periodic driven Hamiltonian 
\begin{equation}
    \hat{H} = \hat{H}_0 + \sum_{j=1}^{\infty} (\hat{H}_j e^{ij\omega t } + \hat{H}_{-j} e^{ij\omega t }).
    \label{eq-floquet-generic}
\end{equation}
From these expressions, we can find that although Eqs. \ref{eq-quantum-BEC} and \ref{eq-quantum-BEC-tranform} have the same eigenvalues, they would yield different Floquet Hamiltonian up to the order of $\omega^{-2}$, as we discussed below. This method may be used to realize new interactions for novel physics. 
For example, it has been used to engineer the topological phase transitions \cite{Lindner2011Floquet, Maczewsky2017Observation, Zhou2023Pseudospin,Weitenberg2021Tailoring,Zhang2023Tuning,Adiyatullin2023Topological,Wintersperger2020Realization} and to realize the discrete time crystal \cite{Zhang2017Observation, Kyprianidis2021Observation}.

\subsection{Method Q-I }
\label{sec-method-1}
In this method, we directly access the Hamiltonian of Eq. \ref{eq-quantum-BEC}. 
We can write the whole Hamiltonian as 
\begin{equation}
    \hat{H} = \hat{H}_0 + \hat{H}_{-1}e^{-i\omega t} + \hat{H}_{1}e^{i\omega t},
    \label{eq-method1-ham}
\end{equation}
with $\hat{H}_0 = \Delta(\hat{n}_a- \hat{n}_b) - \nu (\hat{a}^{\dagger}\hat{b} + \hat{b}^{\dagger}\hat{a}) + \frac{\lambda}{2}(\hat{n}_a- \hat{n}_b)^2 - \frac{\lambda}{2}(\hat{n}_a + \hat{n}_b)$ and $\hat{H}_{-1} = H_{1} =\frac{A}{2}(\hat{n}_a- \hat{n}_b)$. By a direct calculation, we find the  
new effective terms as 
\begin{align}
\hat{H}^{(1)}_{\text{F}} &= \sum_j \frac{1}{\omega j} [\hat{H}_{j},\hat{H}_{-j}] = 0, \label{eq-method1-floquet-firstorder} \\
 \hat{H}^{(2)}_{\text{F}}  &= \frac{A^2}{\omega^2} \nu (\hat{a}^{\dagger} \hat{b} + \hat{b}^{\dagger}\hat{a}). \label{eq-method1-floquet-secondorder}
\end{align}
Combining the above results will yield the effective Floquet Hamiltonian in Eq. \ref{eq-Floquet-quantum} with modified parameters as 
\begin{equation}
    \begin{aligned}
        \Delta' = \Delta, \quad \lambda' = \lambda, \quad
        \nu' = (1 - \frac{A^2}{\omega^2})\nu, \quad \hat{V}_{\text{new}} = 0.
    \end{aligned}
    \label{eq-method1-summary}
\end{equation}
In this method, the periodic driving can only weaken the strength of the coupling.
However, it would not change the parameters of $\lambda$ and $\Delta$ at least up to the order of $\omega^{-2}$.
Moreover, according to Eq. \ref{eq-self-trapping-criteria}, we find this Floquet Hamiltonian predicts the criterion for MSTP as
\begin{equation}
    \lambda_c = 2\nu( 1 - \frac{A^2}{\omega^2}),
\end{equation}
when $\Delta = 0$, $a(0) = 1$, and $b(0) = 0$. This is a good approximation when $|A| < \omega$; see the expression $J_0$ below.  

\subsection{Method Q-II }
\label{sec-method-2}

The Hamiltonian in Eq. \ref{eq-quantum-BEC-tranform} can be divided into 
\begin{equation}
    \hat{H}'
    = \hat{H}_0 + \hat{H}_1(t),
    \label{eq-method-2}
\end{equation}
with 
\begin{align}
\hat{H}_0 &= \Delta (\hat{n}_a - \hat{n}_b) -\nu J_0( \hat{a}^{\dagger}\hat{b} + \mathrm{H.c.}) \nonumber \\
& + \frac{\lambda}{2}[(\hat{n}_a - \hat{n}_b)^2 - (\hat{n}_a + \hat{n}_b) ], \label{eq-method-2-zero} \\
\hat{H}_1 (t) &= \sum_{j = 1}^{\infty}[ \nu J_{j}\hat{a}^{\dagger}\hat{b} +\nu J_{-j}\hat{b}^{\dagger}\hat{a}]e^{ij\omega t } + \mathrm{H.c.} \nonumber \\
& = \sum_{j = 1}^{\infty} H_{j }e^{ij\omega t } + \mathrm{H.c.}.
\end{align}
Here, we have employed the Jacobi-Anger expansion $e^{iz\sin(x)} = \sum_{j=-\infty}^{\infty}  J_j(z)e^{i j x} $ with $J_j(z)$ being the Bessel function of the first kind. Hereafter for convenience we define $J_j = J_j (\frac{2A}{\omega})$.
It can be found that the time-averaged Hamiltonian $\hat{H}_{0}$ can capture the physics predicted by the method in Sec. \ref{sec-method-1}, since $J_0 \sim 1 - A^2/\omega^2$, which is a good approximation when $|A| < \omega$. 
The resulting Floquet Hamiltonian considering the effect from $\hat{H}_1 (t)$ is give by 
\begin{align}
\hat{H}^{(1)}_{\text{F}} &= 0, \\
\hat{H}^{(2)}_{\text{F}} & = - \frac{4 \nu^2 B \Delta}{\omega^2} (\hat{n}_a - \hat{n}_b) - \frac{2 \nu^3 D}{\omega^2} (\hat{a}^{\dagger}\hat{b} + \hat{b}^{\dagger} \hat{a} ) \nonumber \\
&- \frac{4 \nu^2 B \lambda}{\omega^2} [(\hat{n}_a - \hat{n}_b)^{2}-(\hat{n}_a + \hat{n}_b)] \nonumber \\
&-\frac{4\nu^2 C \lambda}{\omega^2} (\hat{a}^{\dagger}\hat{a}^{\dagger}\hat{b}\hat{b}+\hat{b}^{\dagger}\hat{b}^{\dagger}\hat{a}\hat{a}) + \frac{8 \nu^2 B \lambda}{\omega^2} \hat{a}^{\dagger}\hat{a} \hat{b}^{\dagger} \hat{b}. \label{eq-method-2-floquet-secon}
\end{align}
with $B = \sum_{j=1}^{\infty} \frac{1}{j^2}J_j J_j$, $C = \sum_{j=1}^{\infty} \frac{1}{j^2}J_j J_{-j}$, and $ D = (C-B) J_0 + \sum_{jl}\frac{2}{3 j l} [(-1)^{l} - (-1)^{j+l}](J_{-l}J_{l+j}J_{-j} -J_{-l}J_{l-j}J_{j})$. 
As a result, in Eq. \ref{eq-Floquet-quantum}, we have
    \begin{align}
        \Delta' &= (1 - \frac{4\nu^2B}{\omega^2})\Delta, \\
        \lambda' &= (1 - \frac{8\nu^2B}{\omega^2})\lambda, \\
        \nu' &= (J_0 + \frac{2\nu^2D}{\omega^2})\nu, \\
        \hat{V}_{\text{new}} = -\frac{4\nu^2 C \lambda}{\omega^2} & [\hat{a}^{\dagger}\hat{a}^{\dagger}\hat{b}\hat{b} + \mathrm{H.c.}]  + \frac{8 \nu^2 B \lambda}{\omega^2} \hat{a}^{\dagger} \hat{a} \hat{b}^{\dagger} \hat{b}.
\label{eq-method-2-new}
\end{align}
This approximation generates a new term for co-tunneling between the two wells \cite{Folling2007Direct}, which still preserves the total particle number $\hat{n}_a + \hat{n}_b$.
The appearance of this term can be understood from the two-photon process of the periodic-driven Hamiltonian in Eq. \ref{eq-method-2} and it would be the dominant tunneling mechanism when the direct process is suppressed \cite{Lignier2007Dynamical}.
For instance, when we set $2 A/\omega $ at the zero points of the Bessel function $J_0$, it is expected the tunneling between the two wells is completely suppressed from the zero-order Floquet Hamiltonian in Eq. \ref{eq-method-2-zero}.
However, the tunneling of the whole system persists due to the co-tunneling process.

From this result, we find that in this effective model, all parameters are changed. In particular, new interaction 
$V_{\text{new}}$ emerges, which is completely different from the result in Sec. \ref{sec-method-1}. 
As discussed above, this term may lead to observable features. 
When $A < \omega$, following the criterion presented in Appendix \ref{appendix-pendulum}, we find a similar criterion for 
MSTP at 
\begin{equation}
    \lambda_c = 2 \nu \frac{\omega^2 J_0 + 2\nu^2D}{\omega^2 - 12\nu^2 B + 4\nu^2 C}.
    \label{eq-criteria-method-2}
\end{equation}
When $|\omega| \gg |A|$ and $|\nu|$, the Bessel functions have the following asymptotic forms $    J_j = J_j(\frac{2A}{\omega}) \sim \frac{1}{\Gamma(j+1)}(\frac{A}{\omega})^{j}$, 
with $j\geq 1$, from which we expect
\begin{align}
    B \sim  \frac{A^2}{\omega^2}, \quad C \sim -\frac{A^2}{\omega^2}, \quad D \sim -2\frac{A^2}{\omega^2} J_0.
\end{align}
Thus, the criterion for MSTP becomes $\lambda_c \sim 2\nu (1 - \frac{A^2}{\omega^2})$, which is consistent with method Q-I. This result demonstrate that to the same order of $\omega^{-2}$, the different approximations can yield different effective Hamiltonians. The similar features will be observed in the mean field models. 

\section{Floquet analysis based on mean field model}
\label{sec-effective-ham-classic}

The Floquet Hamiltonian can also be obtained from the mean field Hamiltonian and its associated equation of motion. In this way, the 
operators are repalced by $\mathbb{C}$-numbers. 
The general equation of motion can be written as \cite{Wu2000Nonlinear, Choi1999BoseEinstein, Ananikian2006GrossPitaevskii, Ruostekoski1998BoseEinstein, Lyu2020Floquet}
\begin{align}
    i \frac{d a}{dt} = \gamma(t) a + \lambda (|a|^2 - |b|^2) a - \nu(t) b, \\
     i \frac{d b}{dt} = -\gamma(t) b - \lambda (|a|^2 - |b|^2) b -\nu(t) a. \label{eq-mean-field}
\end{align}
The equation of motion of $a^*$ and $b^*$ can be obtained by making a complex conjugate to the above equations, thus is not shown. The same preterition will be made below. When $\gamma(t) = \Delta + A\cos(\omega t )$ and $\nu(t) = \nu$, the equation captures the dynamics of Eq. \ref{eq-quantum-BEC}. 
When $\gamma(t) = \Delta$ and $\nu(t) = \nu e^{i\frac{2A}{\omega}\sin(\omega t )}$, the equation captures the dynamics of Eq. \ref{eq-quantum-BEC-tranform}.
In the following, we use multiple scale analysis to derivate the Floquet Hamiltonian and term the two methods as methods C-I and C-II.
In the last part of this section, we have also presented a more simple approach for the Floquet Hamiltonian, termed method C-III.
A general treatment of this approach can be found in the mechanics textbook \cite{Landau1976Mechanics}. 

\subsection{Method C-I}
\label{sec-method-3}

Following the approach developed by Rahav \textit{et. al} \cite{Rahav2003Effective}, we separate the dynamic variables into slow and fast modes as
\begin{align}
    a &= a_0 + \xi_{a} = a_0 + \sum_{j = 1}^{\infty}\frac{1}{\omega^j} \xi_{a,j},\\
    b &= b_0 + \xi_{b} = b_0 + \sum_{j = 1}^{\infty}\frac{1}{\omega^j}\xi_{b,j}, 
\end{align}
with $\xi_a$ and $\xi_b$ (neglecting the $j$ index) being time-dependent as a function of $\tau = \omega t$ for fast modes, $a_0$ and $b_0$ for the slow modes, respectively. 
Obviously, $\xi_a$ and $\xi_b$ also depend on $a_0$ and $b_0$; see discussion of Karpitz inverted pendulum in Chap 5 of Ref. \cite{Landau1976Mechanics}. Recently, a quantum version Karpitz pendulum was realized with ultracold atoms in Ref. \cite{Jiang2023Karpitz}.
We can assume that $\int_0^{T}dt \xi_{a} = \int_0^{T}dt \xi_b = 0$, with $T = 2\pi/\omega$ the period. In this way, all the fast oscillating terms are averaged out, leaving only the slow variation terms for the effective dynamics. Then we have
\begin{align}
    \frac{d \xi_{a}}{dt} &= \omega \frac{\partial \xi_{a}}{\partial \tau} + \frac{\partial \xi_{a}}{\partial a_0 }\dot{a}_0 + \frac{\partial \xi_{a}}{\partial b_0} \dot{b}_0 + \frac{\partial \xi_{a}}{\partial a_0^{*} }\dot{a}_0^{*} + \frac{\partial \xi_{a}}{\partial b_0^{*}} \dot{b}_0^{*}, \\
    \frac{d \xi_{b}}{dt} &= \omega \frac{\partial \xi_{b}}{\partial \tau} + \frac{\partial \xi_{b}}{\partial a_0 }\dot{a}_0 + \frac{\partial \xi_{b}}{\partial b_0} \dot{b}_0 +  \frac{\partial \xi_{b}}{\partial a_0^{*} }\dot{a}_0^{*}+ \frac{\partial \xi_{b}}{\partial b_0^{*}} \dot{b}_0^{*}.  
\end{align}
The complete conjugate of these results can give the equation of motion of $\xi_a^*$ and $\xi_b^*$. Then we can solve the nonlinear equation in Eq. \ref{eq-mean-field} according to the order of $\omega^{-1}$.
In this method, we choose $\gamma(t) = \Delta + A\cos(\omega t )$ and $\nu(t) = \nu$. 
In practice, we gather all the terms according to $\omega^{-n}$ and then choose the solution of $\xi_{n+1}$ to cancel the dependence of $\tau$.
The constant terms in the next order $\omega^{-(n+1)}$ give rise to the effective contribution with order $\omega^{-(n+1)}$ for the slow part $a_0$ and $b_0$. 

(1) \textit{Zero order}. We have the following equations
\begin{align}
i(\frac{d a_0}{dt} + \frac{\partial \xi_{a,1}}{\partial \tau} ) = & \gamma(\tau) a_0 + \lambda z a_0 - \nu b_0, \label{eq-mean-field-zero-a}\\
i(\frac{d b_0}{dt} +  \frac{\partial \xi_{b,1}}{\partial \tau})= & -\gamma(\tau) b_0 - \lambda z b_0 - \nu a_0,  
\label{eq-mean-field-zero-b}
\end{align}
with $z=|a_0|^2 - |b_0|^2$.
To cancel the dependence of $\tau$, we chose
$i \frac{\partial \xi_{a,1}}{\partial \tau} = A\cos(\tau)a_0$ and $i \frac{\partial \xi_{b,1}}{\partial \tau} = -A\cos(\tau)b_0$, which yield the solutions
\begin{equation}
\begin{aligned}
     \xi_{a,1}(\tau) =  -i A \sin(\tau) a_0, \quad \xi_{b,1}(\tau) = i A \sin(\tau) b_0.
\end{aligned}
\label{eq-mean-field-solution-first}
\end{equation}
Substitute these solutions back to Eqs. \ref{eq-mean-field-zero-a} and \ref{eq-mean-field-zero-b}, we got the zero-order effective equation for $a_0$ and $b_0$, which consistent with time-averaged mean field Hamiltonian

(2) \textit{First order}.
The equations in this order are
\begin{align}
   i\frac{\partial \xi_{a,2}}{\partial \tau} + i\dot{a}_0 \frac{\partial \xi_{a,1}}{\partial a_0 } = (\gamma(\tau) + 2\lambda|a_0|^2 - \lambda |b_0|^2)\xi_{a,1} \nonumber \\
   - (\nu + \lambda b_0^{*}a_0 )\xi_{b,1} + \lambda (a_0)^2 \xi_{a,1}^{*} -\lambda a_0 b_0 \xi_{b,1}^{*}, \\
 i\frac{\partial \xi_{b,2}}{\partial \tau} + i\dot{b}_0 \frac{\partial \xi_{b,1}}{\partial b_0 } = -(\gamma(\tau) + \lambda|a_0|^2 - 2 \lambda |b_0|^2)\xi_{b,1} \nonumber \\
 - (\nu + \lambda a_0^{*}b_0)\xi_{a,1} + \lambda (b_0)^2 \xi_{b,1}^{*} -\lambda a_0 b_0 \xi_{a,1}^{*}.   
\end{align}
It can be found that the solution in this order should be
\begin{align}
     \xi_{a,2}(\tau) &=   2 A \nu  \cos(\tau) b_0 -\frac{A^2}{4}\cos(2\tau)a_0, \\
     \xi_{b,2}(\tau) &=  -2 A \nu  \cos(\tau) a_0 + \frac{A^2}{4}\cos(2\tau)b_0.   
\end{align}
In this order, we can not find any constant terms for $a_0$, $b_0$, and their conjugate.
Thus, we find $H_{\text{F}}^{(1)} = 0$.

(3) \textit{Second order}.
In this order, the equations are much longer than the previous equations.
Thus, we focus on the equation for $a$, while the derivation for $b$ is also straightforward.
The equation for $a$ reads as 
\begin{align}
i\frac{\partial \xi_{a,3}}{\partial \tau} &+ i\dot{a}_0 \frac{\partial \xi_{a,2}}{\partial a_0 } + i\dot{b}_0 \frac{\partial \xi_{a,2}}{\partial b_0 } = \nonumber \\
&(\gamma(\tau) + 2\lambda|a_0|^2 - \lambda |b_0|^2)\xi_{a,2} - (\nu + \lambda b_0^{*}a_0 )\xi_{b,2} \nonumber \\
&+ \lambda (a_0)^2 \xi_{a,2}^{*} -\lambda a_0 b_0 \xi_{b,2}^{*} + \lambda a_0^{*} \xi_{a,1}^2 + 2\lambda a_0 |\xi_{a,1}|^2 \nonumber \\
&- \lambda b_0^{*} \xi_{b,1} \xi_{a,1} - \lambda b_0 \xi_{b,1}^{*} \xi_{a,1} - \lambda a_0 \xi_{b,1}^{*} \xi_{b,1}.
\label{eq-mean-field-second}
\end{align}
Gathering all the constant terms, we find the whole effective equation of motion for $a_0$ and $b_0$ up to the order of $\omega^{-2}$ is
\begin{align}
    i\frac{d a_0}{d t} = \Delta a_0 + \lambda' (|a_0|^2 - |b_0|^2)a_0 - \nu' b_0, \\
    i\frac{d b_0}{d t} = - \Delta b_0 - \lambda' (|a_0|^2 - |b_0|^2)b_0 - \nu' a_0,    
\end{align}
with modified parameters in the Floquet Hamiltonian as 
\begin{equation}
    \begin{aligned}
        \Delta' = \Delta,& \quad \lambda' = (1 + \frac{A^2}{2 \omega^2})\lambda, \\
        V_\text{new} = 0, &\quad \nu' = (1 - \frac{A^2}{ \omega^2}) \nu.
    \end{aligned}
\end{equation}
Similar to the results we obtained in Sec. \ref{sec-method-1}, we find the periodic driving suppresses the coupling strength.
However, it also enhances the interaction strength $\lambda$.
Therefore, with this Floquet Hamiltonian, it is expected that the criterion for MSTP become 
\begin{equation}
    \lambda_c = 4 \nu \frac{\omega^2 - A^2}{2\omega^2 + A^2},
    \label{eq-lambdac-cI}
\end{equation}
when $\Delta = 0$, $a(0) = 1$, and $b(0) = 0$. When $|A/\omega| \ll 1$, we have $\lambda_c = 2\nu (1-3A^2/2\omega^2)$. 

\subsection{Method C-II}

Here, we choose $\gamma(t) = \Delta$ and $\nu(t) = \nu e^{i\frac{2A}{\omega}\sin(\omega t )}$.
Then, we employ a similar method that we have used in Sec. \ref{sec-method-3}.
we can obtain the effective equations for $a_0$, $b_0$, $a_0^{*}$, and $b_0^{*}$ order by order.

(1) \textit{Zero order}. The effective equations in this order are
\begin{align}
    i\frac{d a_0}{dt} + i \frac{\partial \xi_{a,1}}{\partial \tau}&= \Delta a_0 + \lambda z a_0 - \nu J_0  b_0 - \mathcal{B}(\tau) b_0,\\
    i\frac{db_0}{dt} + i \frac{\partial \xi_{b,1}}{\partial \tau} & = - \Delta   b_0 - \lambda z b_0 - \nu J_0 a_0 - \mathcal{B}^{*}(\tau) a_0,
\end{align} 
with $\mathcal{B}(\tau) = \nu \sum_{j= 1}^{\infty} [J_{j} e^{ij\tau} + J_{-j}e^{-ij\tau} ]$ and $z$ same to Eq. \ref{eq-mean-field-zero-a}.
We chose the solution of $\xi_{a,1}$ and $\xi_{b,1}$ as
\begin{equation}
\begin{aligned}
\xi_{a,1} = \mathcal{C}(\tau) b_0, & \quad
    \xi_{b,1} = -\mathcal{C}^{*}(\tau) a_0,
\end{aligned}
    \label{eq-method-4-solution-zero}
\end{equation}
with $\mathcal{C}(\tau) = \sum_{j= 
 1}^{\infty} \frac{\nu}{j}(J_{j} e^{ij\tau} - J_{-j} e^{-ij\tau} )$.

(2) \textit{First order}. The equations in this order are
\begin{align}
  & i\frac{\partial \xi_{a,2}}{\partial \tau} + i\frac{ d b_0}{d t} \frac{\partial \xi_{a,1}}{\partial b_0 } = (\Delta + 2\lambda|a_0|^2 - \lambda |b_0|^2)\xi_{a,1} - \nonumber \\
   &( \nu J_0 + \mathcal{B}(\tau) + \lambda b_0^{*}a_0 )\xi_{b,1} + \lambda (a_0)^2 \xi_{a,1}^{*} -\lambda a_0 b_0 \xi_{b,1}^{*}, \\
    & i\frac{\partial \xi_{b,2}}{\partial \tau} + i\frac{ d a_0}{d t} \frac{\partial \xi_{b,1}}{\partial a_0 } = -(\Delta + \lambda|a_0|^2 - 2 \lambda |b_0|^2)\xi_{b,1} - \nonumber \\
    &(\nu J_0 + \mathcal{B}^{*}(\tau) + \lambda a_0^{*}b_0)\xi_{a,1} + \lambda (b_0)^2 \xi_{b,1}^{*} -\lambda a_0 b_0 \xi_{a,1}^{*}.
\end{align}
With these equations, we can find the solutions for $\xi_{a,2}$ and $\xi_{b,2}$ are
\begin{align}
&\xi_{a,2} = -\mathcal{D}(\tau) (2\Delta b_0 + 4 \lambda |a_0|^2 b_0 - 2 \lambda |b_0|^2 b_0 + \nu J_0 a_0 ) \nonumber \\
&+ \mathcal{D}^{*}(\tau)  (\nu J_0 a_0 + 2\lambda a_0^{2} b_0^{*})
+\mathcal{F}(\tau) a_0, \\
&\xi_{b,2} = \mathcal{D}^{*}(\tau) ( 2\Delta a_0 + 2 \lambda |a_0|^2 a_0 - 4 \lambda |b_0|^2 a_0  - \nu J_0 b_0 ) \nonumber \\
&+ \mathcal{D}(\tau)  (\nu J_0 b_0 + 2\lambda b_0^2 a_0^{*})
+ \mathcal{F}^{*}(\tau) b_0,
\end{align}
with $\mathcal{D}(\tau) = \sum_{j=1}^{\infty} \frac{\nu}{j^2}(J_{j} e^{ij\tau} + J_{-j} e^{-ij\tau} )$ and $\mathcal{F}(\tau) = \sum_{j\neq 0, l \neq 0, j \neq l} = \frac{J_{j}J_{l}}{(j-l)l}e^{i(j-l)\tau}$.
Here, $j$ and $l$ run from negative infinity to positive infinity.
The solution for $\xi_{a,1}^{*}$ and $\xi_{b,1}^{*}$ can also be obtained by applying conjugation.

(3) \textit{Second order}. The constant terms in this order will contribute to the effective equations.
The resulting Floquet Hamiltonian at this order is
\begin{eqnarray}
    H_{\text{F}}^{(2)} && = -\frac{4\nu^2B\Delta}{\omega^2}(|a_0|^2 - |b_0|^2) - \frac{3\nu^2 B \lambda}{\omega^2} (|a_0|^2 - |b_0|^2)^2 
    \nonumber \\ 
    &&- \frac{2\nu^3M}{\omega^2} 
    (a_0^{*}b_0 + b_0^{*}\tilde{a}_0) + \frac{8 \nu^2 B \lambda}{\omega^2} |a_0|^2|b_0|^2 \nonumber \\ 
    && - \frac{4 \nu^2 C \lambda}{\omega^2} (a_0^{*}a_0^{*}b_0b_0+b_0^{*}b_0^{*}a_0a_0),
\end{eqnarray}
with $B$ and $C$ being given by Eq. \ref{eq-method-2-floquet-secon}, M being given by $
M = (C-B) J_0 +  \sum_{j,l>0,j\neq l}\frac{1}{jl} (J_{-j} J_{j+l}J_{l}-J_{-j} J_{j-l}J_{-l})$.    
Compared with Eq. \ref{eq-Floquet-mean-field}, we find
\begin{eqnarray} 
    \Delta' &&= (1 - \frac{ 4\nu^2B}{\omega^2})\Delta,  \\ 
    \lambda' &&= (1 - \frac{6\nu^2B}{\omega^2})\lambda, \\ 
    \nu' &&= (J_0 + \frac{2\nu^2M}{\omega^2})\nu,\\
    V_{\text{new}} && = - \frac{4 \nu^2 C \lambda}{\omega^2 } (a^{*}a^{*}bb+b^{*}b^{*}aa) + \frac{8 \nu^2 B \lambda}{\omega^2}  |a|^2|b|^2 . \nonumber  \\ 
    \label{eq-method-4-new}
\end{eqnarray}
In this method, we also find the co-tunneling term as we have discussed in method Q-II.
Similarly, the criterion for MSTP becomes (see Appendix \ref{appendix-pendulum})
\begin{equation}
    \lambda_c = 2\nu \frac{ \omega^2 J_0 + 2 \nu^2M}{ \omega^2-10\nu^2 B + 4\nu^2 C}.
    \label{eq-criteria-method-4}
\end{equation}
Using a similar analysis in method Q-II, we find  $ \lambda_c \sim 2\nu (1 - \frac{A^2}{\omega^2})$, at the limit of $|\omega| \gg |A|$ and $|\nu|$.

%

\subsection{Method C-III}
\label{sec-method-5}

In this method, we use a much simpler but useful approach for the derivation of the Floquet Hamiltonian, following the essential idea in the Karpitz pendulum \cite{Landau1976Mechanics}. We expand $a$ and $b$ into the slow and fast modes up to the order of $\omega^{-2}$ as 
\begin{equation}
a = a_0 + a_1 + a_2 , \quad b = b_0 + b_1 + b_2,
\end{equation}
with $a_j$ and $b_j$ of the $\omega^{-j}$ order.
Using this expansion, we can rewrite the nonlinear equations as 
\begin{align}
   & i\frac{d}{dt}(a_0 + a_1 + a_2) = \Delta (a_0 + a_1 + a_2) + \lambda (|a_0+a_1 + a_2|^2 \nonumber \\ 
   & - |b_0 + b_1 + b_2|^2) (a_0 + a_1 + b_2) - \nu (b_0 + b_1 + b_2) \nonumber \\
   &+ A \cos(\omega t) (a_0 + a_1 + a_2 ), \\
   & i\frac{d}{dt}(b_0 + b_1 + b_2 ) =  -\Delta (b_0 + b_1 + b_2) - \lambda (|a_0+a_1 + a_2|^2 \nonumber \\
    & - |b_0 + b_1 + b_2|^2) (b_0 + b_1 + b_2) - \nu (a_0 + a_1 + a_2) \nonumber \\
    &- A \cos(\omega t) (b_0 + b_1 + b_2).
\end{align}
Similarly, we can gather all terms according to the power of $\omega$.
We have the equations of $a_1$ and $b_1$, as follows 
\begin{equation} 
i\dot{a}_1 = A \cos(\omega t) a_0, \quad i\dot{b}_1 = - A \cos(\omega t) b_0,
\end{equation} 
with solution 
\begin{equation}
a_1 = -i {A \sin(\omega t) \over \omega} a_0, \quad  b_1 = i {A \sin(\omega t) \over \omega} b_0.
\end{equation}
Strictly speaking, the above solution is not accurate if we consider the time-dependence of $a_0$ and $b_0$, making $a_1(\tau, t)$, with $\tau = \omega t$.
In the multiple scale analysis as discussed in methods C-I and C-II, the time-dependence of $a_0$ and $b_0$ has been taken into account. 
Ignoring this time dependence leads to results with much less accuracy; however, it is still instructive and has pedagogical value \cite{Landau1976Mechanics}. 
In this way, there are no constant terms that can be separated in $\omega^{-1}$ order. 
Then we focus on the solution to the order of $\omega^{-2}$, with 
\begin{align}
    a_2 &= \frac{A}{\omega^2}\cos(\omega t )[\Delta a + \lambda (|a_0|^2 - |b_0|^2)a_0 + \nu b_0] \nonumber \\
    & + \frac{A^2}{(2\omega)^2}\cos(2\omega t)a_0, \\ 
    b_2 &= -\frac{A}{\omega^2}\cos(\omega t )[-\Delta b - \lambda (|a_0|^2 - |b_0|^2)b_0 + \nu a_0] \nonumber \\ 
    &+ \frac{A^2}{(2\omega)^2}\cos(2\omega t)b_0.
\end{align}
In this way, the effective term for $a_0$ comes from 
\begin{align}
&\frac{1}{T}\int_0^{T} dt A\cos(\omega t ) a_2(t) \nonumber \\
&= \frac{A^2}{2\omega^2}[\Delta a_0 + \lambda (|a_0|^2 - |b_0|^2)a_0 + \nu b_0 ],\\
& \frac{1}{T}\int_0^{T} dt \lambda(2 a_0|a_1|^2 + a_0^{*}a_1^2 - a_0 |b_1|^2 - b_0^{*}b_1 a_1 - b_0 b_1^{*}a_1 ) \nonumber \\
& =\frac{A^2}{2\omega^2}\lambda (|a_0|^2 - |b_0|^2)a_0,
\end{align}
and the same analysis can be employed to $b_0$.
Combining all these terms, we have the effective equation of motion for $a_0$ and $b_0$ as follows
\begin{align}
 i\frac{d a_0}{dt} = & + \Delta' a_0  +  \lambda'(|a_0|^2 - |b_0|^2)a_0 - \nu' b_0, 
  \\
 i\frac{d b_0}{dt} = & -\Delta' b_0 - \lambda' (|a_0|^2 - |b_0|^2)b_0 - \nu' a_0,
\end{align}
with 
\begin{align}
    \Delta' &= (1 + \frac{A^2}{2\omega^2})\Delta, \\ 
    \lambda' &= (1 + \frac{A^2}{\omega^2})\lambda, \\
    \nu' &= (1 - \frac{A^2}{2\omega^2})\nu, \quad
    V_{\text{new}} = 0.
\end{align}
In this result, the criterion for the MSTP is
\begin{equation}
    \lambda_c =  \nu\frac{2\omega^2 - A^2}{\omega^2 + A^2},
\end{equation}
for $\Delta = 0$, $a(0) = 1$ and $b(0) = 0$. When $A$ is much smaller than $\omega$, it yields $\lambda_c = 2\nu (1-\frac{3A^2}{2\omega^2})$, which is the same as that from Eq. \ref{eq-lambdac-cI}; please see a direct comparison between different approaches in Table \ref{tab-summary}.

\section{Numerical verification}
\label{sec-comparision}

\begin{figure}[htbp]
    \centering
    \includegraphics[width=0.48\textwidth]{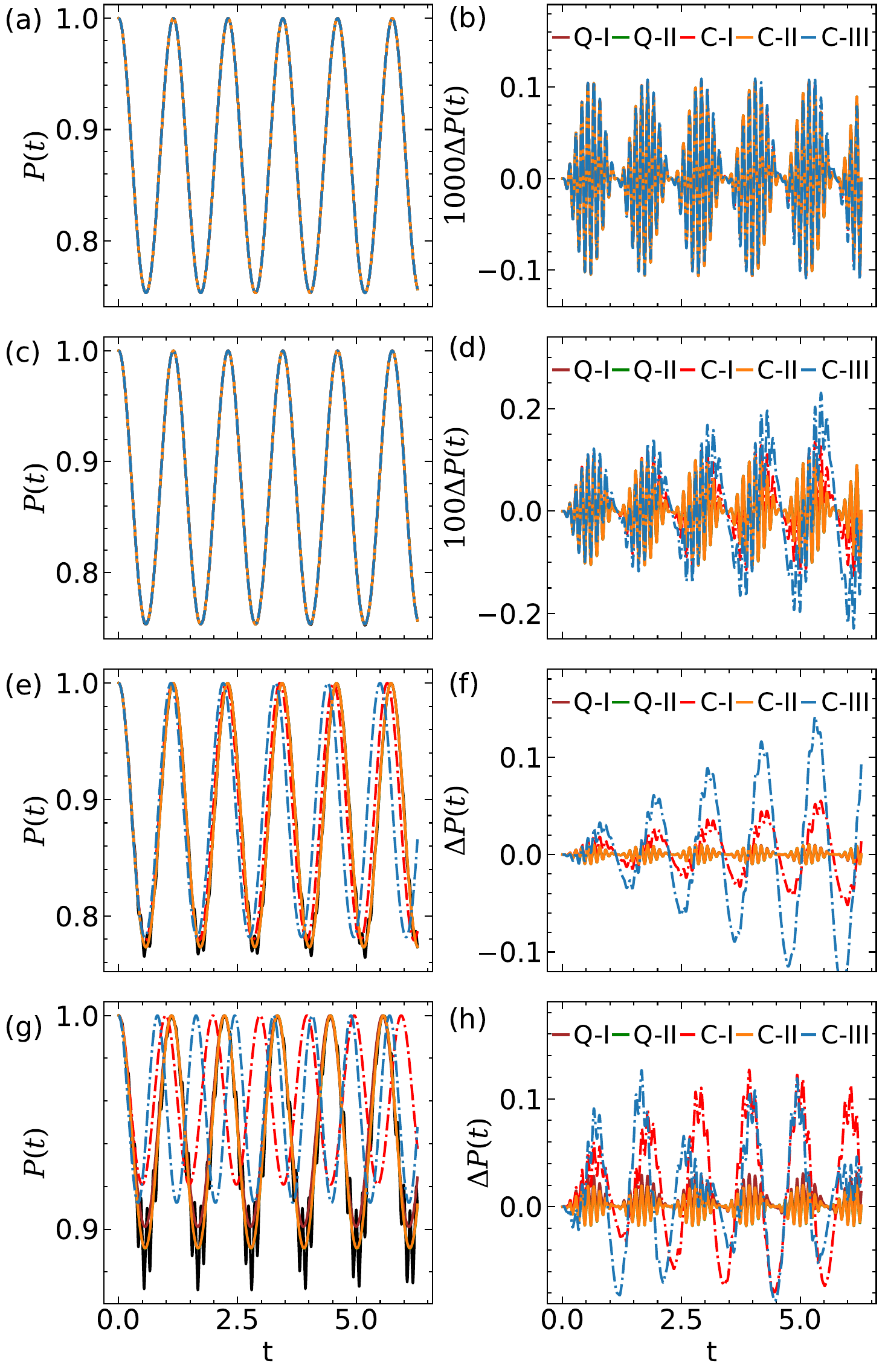}
    \caption{ The left column denotes the population $P(t) = |a|^2 - |b|^2$ against evolution time $t$ using Eq. \ref{eq-mean-field} and other methods.
    The black solid lines are the results obtained by Eq. \ref{eq-mean-field}, which we call $P_0(t)$.
    The right column denotes the 
    difference $\Delta P(t) = P(t) - P_0(t)$ of population dynamics between Eq. \ref{eq-mean-field} and other methods. 
    We have used $\Delta = 1$, $\lambda = 1.9$, $\nu = 1$, and $\omega = 50$ in all figures and we have used $A = 0.1$ in (a) and (b); $A = 1$ in (c) and (d); $A = 10$ in (e) and (f); and $A = 30$ in (g) and (h).
    }
    \label{fig-numerical-verification}
\end{figure}

\begin{figure}[htbp]
    \centering
    \includegraphics[width=0.48\textwidth]{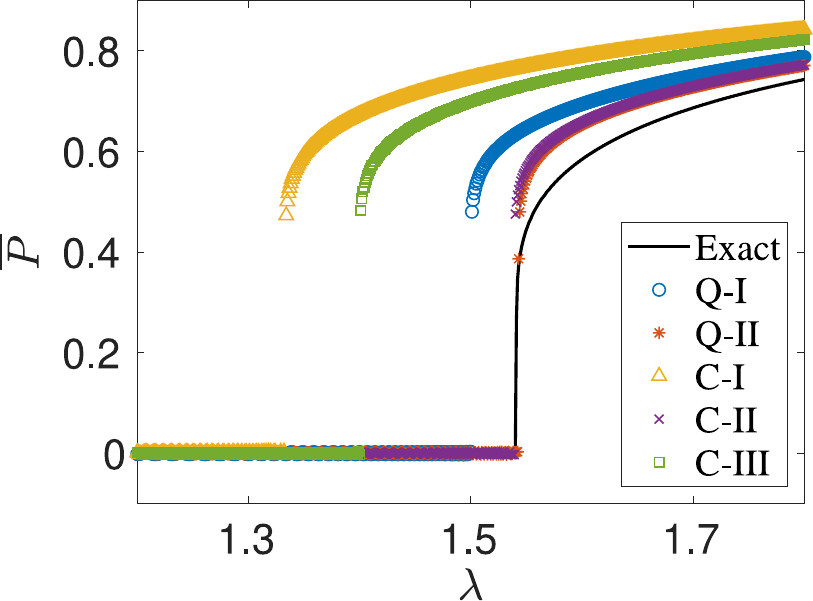}
    \caption{ The time-averaged population difference with different methods. 
    In the calculation of $\overline{P}$, we set $\Delta = 0$, $\nu = 1$, $A = 10$, $\omega = 20$, $a(0) = 1$, $b(0) = 0$, and $M = 500$.
    It is expected $\lambda_c = 1.5$ for method Q-I, $\lambda_c \sim 1.543 $ for method Q-II, $\lambda_c = 4/3 $ for method C-I, $\lambda_c \sim 1.538$ for method C-II, and $\lambda_c \sim 1.4 $ for method C-III. The accurate numerical result gives $\lambda_c = 1.54$.
    }
    \label{fig-criteria}
\end{figure}

With the above five methods and their different Floquet Hamiltonian, it is crucial to ask whether all these Floquet Hamiltonian give rise to a good description of the periodic driven BEC or only one yields the correct physics.  
To answer this question, we numerically solve the mean field equation of motion that is obtained by Eq. \ref{eq-mean-field} and other effective Hamiltonians. 
Then we compare the population dynamics $P(t) = |a|^2 - |b|^2$ from these methods in Fig. \ref{fig-numerical-verification}.
It is found that all methods give rise to an excellent agreement with the exact dynamics in the short time for $A = 0.1$ and $A = 1$ with $ \Delta P(t) \sim 10^{-4}-10^{-3}$.
However, when $A = 10$ and $A = 30$, the difference between the dynamics in Eq. \ref{eq-mean-field} and other methods becomes much larger.
Of all the methods, we find the Q-II and C-II are the most accurate, whose $\Delta P \sim 10^{-2}-10^{-1}$ even for strong periodic driving.
However, methods C-I and C-III, where interaction strength is enhanced by periodic driving, give rise to the worst approximation.
This can also be understood by the example we discussed in Appendix \ref{appendix-exact}.

Then we should verify our prediction for the criterion of the MSTP based on these different methods.
To identify the oscillating phase and MSTP, we use the time average of the population 
\begin{equation}
    \overline{P} = \frac{1}{M T} \int_0^{M T} dt P(t),
\end{equation}
with $T$ the periodicity.
It approaches zero for the oscillating phase while its value will remain finite for the MSTP.
We present the results in Fig. \ref{fig-criteria}.
It can be found that $\overline{P}$ exhibits a transition from zero to finite for all methods.
However, the transition point is different for each method.
It is found that the transition point predicted by methods Q-II and C-II is much closer to the accurate numerical results while the other methods predict much smaller values.
This may be understood from the Bessel function in Eq. \ref{eq-quantum-BEC-tranform}, which includes the contribution higher than $\omega^{-2}$. 
On the other hand, we can find the quantum method Q-I is more accurate than C-I.
This difference should come from the nonlinear feature of method C-I.
Among all methods, method C-III works the poorest in all methods due to the classical approximation and the lack of time-dependent terms in $a_0$ and $b_0$.
This result may be observed in experiments with increasing of resolution in experiments in the future. 

We present several remarks on the above results. (I) The most striking feature of these different methods is that $\lambda'$, $\Delta'$, and $\nu'$ can be unchanged, increased, or decreased, depending strongly on the approximations employed.
However, all of them seem to be reasonable and accurate when $\omega$ is large enough.
(II) In all the classical methods, the variable is divided into fast and slow modes. Thus, the normalization of $|a|^2+|b|^2 = 1$ is not guaranteed in the multiple scale analysis.
In contrast, the Hamiltonian is transformed according to unitary transformation in quantum methods. 
This may account for the inaccuracy of the classical methods when the frequency of the driving becomes small. (III) On the other hand, quantum methods also have their problems. From the quantum Hamiltonian in Eq. \ref{eq-floquet-generic} and perturbative expansion of the Floquet Hamiltonian in Eq. \ref{eq-floquet-quantum-expression}, we can find the Floquet Hamiltonian is highly sensitive to the expression of $\hat{H}_0$ and $\hat{H}_j$. 
Thus the unitary transformation would affect $\hat{H}_0$ and $\hat{H}_j$, leading to different Floquet Hamiltonian in the presence of truncation. In this way, some of the approximations will yield much better results than the other methods. Thus, the choice of the $\hat{H}_0$ and $\hat{H}_j$ by a proper unitary transformation is a rather tricky issue in practice. In Appendix \ref{appendix-exact}, we show that Q-II and C-II based on the unitary transformation will yield the most accurate result. 

\section{Conclusion and discussion}
\label{sec-conclusion}

In this work, we have systematically revisited the Floquet Hamiltonian of the BEC in a double well with periodic driving using five different methods, all of which have solid theoretical grounds and seem to be reasonable.
In all methods, the Floquet Hamiltonian and associated mean field dynamics are discussed up to the order of $\omega^{-2}$.
In method Q-I, we find that periodic driving can only weaken the coupling, while it can not affect other parameters.
In method Q-II, all parameters are modified by periodic driving and we find that both the coupling and the interaction strength are weakened.
Moreover, new interaction terms for co-tunneling, which still preserves the total number of particles, emerge from the second-order process. 
In methods C-I, C-II, and C-III, we use multiple scale analysis methods to obtain the Floquet Hamiltonian.
It is found that the terms, coming from the nonlinear interaction (i.e. $|\xi_{a,1}|^2$, $|\xi_{b,1}|^2$, \text{etc.}), also contribute to the Floquet Hamiltonian.
In method C-III, we have employed the same method used in Karpitz pendulum \cite{Landau1976Mechanics} to derive the Floquet Hamiltonian, yielding results that all parameters are modified.
Finally, we compare all these results on equal footing from the dynamics of the condensate and the associated macroscopic self-trapping process.
We find that while all methods can give a good approximation to the physics in the high frequency limit, significant and observable differences can emerge when the strength of the driving becomes comparable with the modulating frequency. 
It is found that the quantum methods give much better approximation than the classical methods.
Especially, the last method (C-III) yields the poorest approximation to all physics. This justifies that the Floquet Hamiltonian based on Eq. \ref{eq-quantum-BEC-tranform} is more accurate than that based on Eq. \ref{eq-quantum-BEC}. 

These results may indicate that the analysis based on the Floquet Hamiltonian may contain some subjectivity.
If the calculation is performed to infinity orders, one may expect that different starting points may yield the same conclusion.
However, due to the presence of truncation, the different methods can yield different effective Floquet Hamiltonians.
More strikingly, the parameters even get an opposite modification in different methods.
Since all these results give rise to reasonable descriptions at the high frequency limit, it is hard for us to claim that periodic driving can strengthen or weaken some particular parameters.
A similar problem also arise in the other periodic-driven systems.
Therefore, we have to be careful in employing the Floquet Hamiltonian approach to analyze the modulated systems in the future if high precision measurement can be achieved, in which the validity of these different approaches from different approximations can be readily verified using various simulation platforms.
In the current work, we consider the spinless models; and in the presence of spin degrees of freedom, the Floquet engineering can be used to realize new interactions that do not exist in the static Hamiltonian \cite{Jotzu2014Experimental, Eckardt2017Colloquium, Hainaut2018Controlling, Oka2019Floquet, Zhou2023Floquet, Zhao2023Floquet-tailored}, thus the modification of the parameters in the Hamiltonian and the emergence of new terms $\hat{V}_\text{new}$ from $\hat{H}_\text{F}^{(n)}$ is reasonable and meaningful.
In the low modulating limit, some of the above approaches may become invalid, yielding new band structures \cite{Lyu2020Floquet,Vogl2020Effective}. This is a common problem in all driven systems based on various methods \cite{Goldman2014Periodically, Golddman2014Light,Vajna2018Replica, Venkatraman2022Static, Vogel2019Flow}, and the question that which method is more accurate is then important and needs to be concerned in much more careful in future. 

\begin{acknowledgments}
This work is supported by the Strategic Priority Research Program of the Chinese Academy of Sciences with grant No. XDB0500000, and the
Innovation Program for Quantum Science and Technology (No. 2021ZD0301200 and No. 2021ZD0301500).
\end{acknowledgments}

\appendix

\section{The criterion for MSTP}
\label{appendix-pendulum}

In this appendix, we present a complementary derivation for the criterion of MSTP.
It is found that all the mean field Floquet Hamiltonian in this manuscript can be summarized as
\begin{equation}
\begin{aligned}
    H &= \Delta (|a|^2 - |b|^2) - \nu (a^{*}b + b^{*}a) + \lambda_2 |a|^2|b|^2 \\
    &+ \frac{\lambda}{2}(|a|^2 - |b|^2)^2  - \lambda_1 (a^{*}a^{*}bb  + b^{*}b^{*}aa). 
\end{aligned}
\end{equation}
It can be mapped to the following pendulum model
\begin{equation}
    \begin{aligned}
    H_{\text{NP}}  &= \frac{2\lambda-\lambda_2}{4}z^2 + \Delta z - \nu \sqrt{1-z^2}\cos(\phi) \\
    &- \frac{\lambda_1}{2}(1-z^2)\cos(2\phi). 
    \end{aligned}
\end{equation}
When $\lambda_1 = \lambda_2 = 0$, this model reduces to the model in the main text.
Because this model is not chaos, the particle can only move along the constant energy surface.
In the following, we set $\Delta = 0$ while the discussion with $\Delta \neq 0$ is also straightforward.
When $z = 0$, we find 
\begin{equation}
    E_{\text{NP}} = -\nu \cos(\phi) - \frac{\lambda_1 }{2}\cos(2\phi).
\end{equation}
In this work, we mainly focus on the high frequency limit, where the value of $\lambda_1$ is much smaller than $\nu$.
Thus, the maximum of $E$ is at $\phi = \pi$ with $\max(E) = \nu - \frac{\lambda_1}{2}$ and the criterion for MSTP is 
\begin{equation}
\begin{aligned}
   &\frac{2\lambda-\lambda_2}{4}z(0)^2  - \nu \sqrt{1-z(0)^2}\cos(\phi(0)) \\
    &- \frac{\lambda_1}{2}(1-z(0)^2)\cos(2\phi(0))  > \nu - \frac{\lambda_1}{2}
\end{aligned}
\label{eq-criteria}
\end{equation}
Substitute the expressions of $\lambda'$, $\lambda_1$, $\nu'$, and $\lambda_2$ into the above equation will yields $\lambda_c$, which is used in Eq. \ref{eq-criteria-method-4}. We see that in method Q-II, we also have the similar new term (in Eq. \ref{eq-method-2-new}), with the same constant $B$ and $C$. Yet, the parameters of $\nu'$, $\lambda'$ and $\nu'$ in these two approaches (Q-II and C-II) are slightly different. 

\section{Exact solution}
\label{appendix-exact}

For the sake of self-contained, here we present an analysis of the nonlinear model without tunneling (set $\nu = 0$).
In this case, the two equations are totally
decoupled. By a proper phase transformation $a \rightarrow a \exp(iK(t))$ and $b \rightarrow b \exp(-iK(t))$, where $\dot{K} = \Delta + A \cos(\omega t)$, the above two equation becomes 
\begin{equation}
i\dot{a} = \lambda (|a|^2 - |b|^2), \quad  i \dot{b} = - \lambda (|a|^2 - |b|^2). 
\end{equation}
This result is exact, from which we find $i\frac{d}{dt}(a + b) = 0$.
If we use the pendulum analogy, it can be found that
\begin{align}
    \frac{d}{d t} z = 0, \quad \frac{d}{d t} \phi = \lambda z.
\end{align}
The solution is $z = c$, and $\phi = \lambda c t$, with $c$ being a constant.  We find that this solution is fulfilled only by methods Q-   II, and C-II, which are realized by first performing the unitary transformation.

\bibliography{ref}

\end{document}